\documentclass[preprint,pre,floats,aps,amsmath,amssymb]{revtex4}

\usepackage{graphicx}
\usepackage{bm}
\usepackage{comment}

\setcitestyle{super}

\begin{document}

\title{Arginine-Phosphate Salt Bridges Between Histones	 and DNA:	 Intermolecular Actuators that Control Nucleosome Architecture}
\author{Tahir I. Yusufaly$^{1*}$, Yun Li$^{2}$, Gautam Singh$^{3}$ and Wilma K. Olson$^{3}$}
\affiliation{[1] Rutgers, the State University of New Jersey, Department of Physics and Astronomy, Piscataway, NJ, USA, 08854 \\
                   [2] Delaware Valley College, Department of Chemistry and Biochemistry, Doylestown, PA, USA, 18901 \\
		[3] Rutgers, the State University of New Jersey, Department of Chemistry and Chemical Biology and BioMaPS Institute for Quantitative Biology, Piscataway, NJ, USA, 08854 \\ 
		* Corresponding Author. Email: yusufaly@physics.rutgers.edu}
\date{\today}

\begin{abstract}
Structural bioinformatics and van der Waals density functional theory are combined to investigate the mechanochemical impact of a major class of histone-DNA interactions, namely the formation of salt bridges between arginine residues in histones and phosphate groups on the DNA backbone. Principal component analysis reveals that the configurational fluctuations of the sugar-phosphate backbone display sequence-specific directionality and variability, and clustering of nucleosomal crystal structures identifies two major salt-bridge configurations: a monodentate form in which the arginine end-group guanidinium only forms one hydrogen bond with the phosphate, and a bidentate form in which it forms two. Density functional theory calculations highlight that the combination of sequence, denticity, and salt-bridge positioning enable the histones to apply a tunable mechanochemical stress to the DNA via precise and specific activation of backbone deformations. The results suggest that selection for specific placements of van der Waals contacts, with high-precision control of the spatial distribution of intermolecular forces, may serve as an underlying evolutionary design principle for the structure and function of nucleosomes, a conjecture that is corroborated by previous experimental studies.
\end{abstract}

\maketitle

\linespread{1.0}

\section{Introduction}
The discovery of the double-helical structure of DNA \cite{WatsonCrick} established that genetic information is encoded in the molecular sequence of base-paired nucleotides constituting an organism's genome. This information is transduced into an observable set of characteristics, or phenotype, via the central tenet of molecular biology: gene sequences of DNA are transcribed into complementary RNA sequences, which are subsequently translated into functional proteins.  Mutations in DNA provide genetic variability, and Darwinian evolution acts on the resulting diversity of phenotypes, selecting for traits that maximize evolutionary fitness. 

However, modifications of base sequences are not the only source of phenotypic variability. There exists an additional set of modifications termed the epigenetic code, which modify an organism's hereditary information while leaving the genomic sequence intact \cite{Turner07}. While epigenetic regulation occurs at all levels of gene expression, one of the most prominent mechanisms is at the level of control of transcription. In eukaryotes, this occurs via the dynamic remodeling of the structure of chromatin, the bundled assembly of DNA and histones, the structural proteins that package and organize the genomic material. This remodeling controls the expression of specific genes, by selectively blocking or enabling the binding of transcription factors to particular regions of the genome \cite{ChenDent2014}. 

\subsection{The Rise of van der Waals Density Functional Theory}
The importance of the mechanical manipulation of DNA for the control of gene expression has led to the emergence of single-molecule biophysical \cite{Bustamante} experiments that directly probe the molecular machinery operating on DNA at a nanoscale level. However, accurate quantum-mechanical modeling and simulation of these systems is relatively less mature. In particular, while first-principles calculations of `hard' matter have sufficiently advanced to allow the predictive, atomic-level design of new materials before they are synthesized in the laboratory \cite{MRS2006}, they have not been similarly applied to the `soft' biomolecular machinery in the cell. Historically, the key reason for this dearth of activity was the inability of traditional Kohn-Sham Density Functional Theory (KS-DFT) \cite{HK64,KS65} to account for the nonlocal London dispersion forces that are ubiquitous in soft matter.

The recent development of van der Waals density functional theory \cite{Dion04, Lee10} (vdW-DFT) has remedied this situation, expanding the realm of DFT to soft and biological materials. Subsequent applications of vdW-DFT have yielded novel atomistic insight into biologically important mechanochemical processes in DNA. Cooper et. al \cite{Cooper08} studied the hydrogen bonding between base pairs and stacking interactions between nearest-neighbor nucleic acid base-pair steps, and illustrated the role of these interactions in determining sequence-specific elasticity. A follow-up study \cite{Yusufaly13} investigated the 5-methylation of cytosines in 5$^{'}$-CG-3$^{'}$ : 5$^{'}$-CG-3$^{'}$ base-pair steps, an epigenetic modification that is thought to trigger the protein-assisted compaction of chromatin \cite{BirdWolfe99}.

\subsection{The Importance of Histone-DNA Interactions }
Chemical changes to the nucleobases, however, are only a small piece of the elaborate epigenetic machinery controlling DNA structure. Further advances in the usefulness of density functional theory for molecular biophysics will inevitably require expanding its application to a more diverse group of biomolecular processes. In the context of the regulation of chromatin architecture, the interactions between histones and DNA are a timely example of an important class of processes that are ripe for investigation.

Recent theoretical and experimental work \cite{Clauvelin1, Clauvelin2} has highlighted the role of histones, the structural proteins that package chromatin, in mediating long-range communication between regulatory elements in the genome. The physical mechanism behind this signaling is the controlled manipulation of DNA elasticity at specific genomic sites. This is accomplished through a complex interplay of direct and water-mediated protein-DNA interactions \cite{Luger1, Luger2}.

Over thirty years ago, Mirzabekov and Rich \cite{Mirzabekov} suggested that histone-DNA interactions control DNA flexibility in chromatin via neutralization of the sugar-phosphate backbone by cationic amino acids. This has inspired several experimental investigations into the electrostatic mechanisms of protein-induced DNA bending. These studies have verified that this counterion condensation, long assumed to be independent of sequence, is indeed a major contributing factor to DNA deformability \cite{StraussReview, Cherstvy1}.

However, in recent years \cite{Marcovitz, Honig}, it has become apparent that this supposedly sequence-independent electrostatic neutralization is not the only significant mode of interaction between cationic amino acids and the polyelectrolyte backbone. Looking beyond simply electrostatic binding, there also exist several additional non-covalent interactions, including hydrogen bonding between amino acids and phosphate groups, cation-$\pi$ interactions between positively charged amino acids and deoxyribose sugars, and van der Waals forces \cite{StructBook}. These additional forces allow for the control of chemical architectures at a higher level of precision.

\subsubsection*{Arginine-Phosphate Salt Bridges}

\begin{figure}[h]  
\includegraphics[scale=.6]{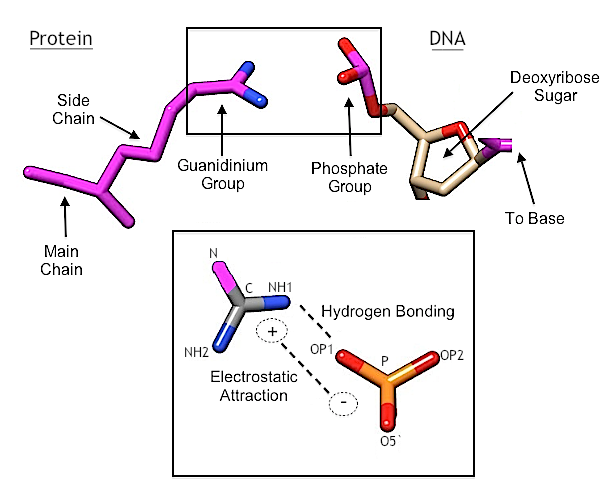}
\caption{In a salt bridge between a histone protein and DNA, the guanidinium side-chain group of the amino acid arginine (top left) binds to the phosphate group of the DNA sugar-phosphate backbone (top right). This is done through a combination of: 1) Electrostatic attraction between the negatively-charged phosphate and positively-charged guanidinium, and 2) Hydrogen bonds between the two end-group nitrogens in guanidinium, labelled NH1 and NH2, and the two side-group oxygens on the phosphate, labelled OP1 and OP2. C and N label the carbon and the non-end group nitrogen on the guanidinium, respectively. O5$^{'}$ is an oxygen connecting to the main chain of the sugar-phosphate backbone. Image created with Pymol \cite{Pymol}.}
\label{saltbridge}
\end{figure}

A relevant example of the interplay between these different molecular forces is the salt bridge between the side-chain guanidinium cation of arginine and the phosphate group of the DNA backbone, as illustrated in Figure \ref{saltbridge}. This salt bridge is one of the most common mechanisms by which histones bind to DNA \cite{Luger1, Luger2}. It consists of a combination of electrostatic attraction between the charged molecular entities and hydrogen bonds of the guanidinium nitrogens to the phosphate group oxygens.

\subsection{The Present Work}

While there have been previous quantum-mechanical studies of the energetics of the sugar-phosphate backbone \cite{Sponer1, Sponer2}, including some work on arginine-phosphate interactions \cite{Frigyesa}, such studies have not yet been attempted using the most recent vdW-DFT methods. With these guidelines in mind, the current work presents a novel investigation into the effects of guanidinium-phosphate salt bridges on the local conformational elasticity of the DNA sugar-phosphate backbone.

Useful application of first-principles calculations to biophysics, however, crucially requires that they not become divorced from the biological context of the problem at hand. In this regard, it is valuable to bridge the traditional gap between the electronic structure theory and structural bioinformatics communities. The latter can help with the judicious selection of biologically relevant molecular configurations to subject to more detailed atomistic modeling. In particular, principal component analysis (PCA) of a statistical ensemble of experimental crystal structures reduces the intractably large phase space of possible molecular deformations to an `essential subspace' of slow modes, or low-frequency collective motions most associated with biological functionality \cite{Yang08, Hub09}. Density functional theory can then provide quantitative information regarding how these functional motions are influenced by specific biochemical perturbations. Electronic structure calculations thus serve as a complement to single-molecule experiments, allowing a microscopic view of the detailed mechanochemical machinery operating within living cells. 

The paper is organized as follows: after an introduction to the basic modeling setup of the problem, the relevant bioinformatics analysis and electronic structure procedures are described. The main results of the work are then presented and discussed. The principal components of fluctuation of the sugar-phosphate backbone are observed to encode sequence information, and DFT calculations illustrate that salt bridges non-covalently interact with the sugar-phosphate backbone in a complex, multi-faceted manner, enabling precision-controlled activation of various backbone deformations. The results have implications for how specific local histone-DNA interactions and positions can stabilize and control more global, long-range elastic landscapes, an effect that is important for nucleosome positioning. As an important corollary, the evolutionary selection for precisely controlled nucleosome positioning in living organisms implies corresponding selection for the precise spatial distribution of these sequence-specific contacts, an effect that is experimentally supported by further analysis of nucleosomal crystal structures and previous molecular biology experiments.

\section{Modeling Setup}
One of the first tasks in modeling arginine-DNA interactions is the selection of an appropriate `model complex' that is a realistic representation of the actual salt bridge and is sufficiently simple to allow for detailed statistical analysis and atomistic calculations. Such a complex should isolate the specific local effects of guanidinium-phosphate hydrogen bonding and electrostatics on DNA deformability.

\begin{figure}[h]  
\includegraphics[scale=.3]{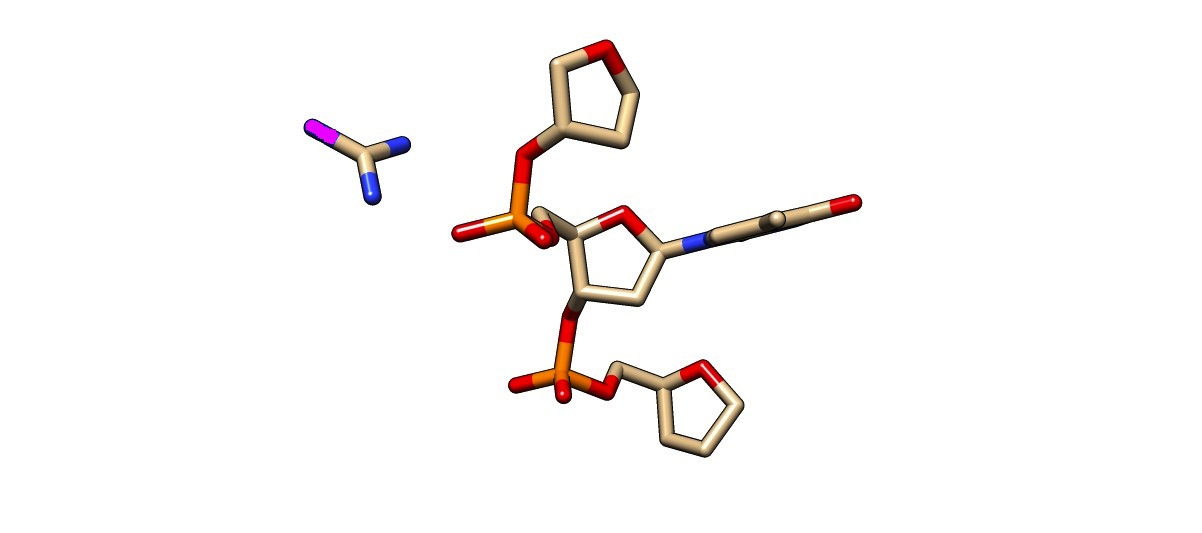}
\caption{The model complex selected for this study consists of a guanidinium cation representative of the end-group of the arginine residues, and a collection of three deoxyribose sugars connected by two intermediate phosphate backbone linkages. Carbon atoms are colored beige, oxygen atoms red, phosphorus atoms orange, and all nitrogens blue except for the non-end group nitrogen of the guanidinium, which is colored purple. Hydrogen atoms are not illustrated for clarity. Image created with Pymol.}
\label{ModelComplex}
\end{figure}

The model complex chosen for this study is illustrated in Figure \ref{ModelComplex}. It strips off all atoms of the arginine amino acid except for the end-group guanidinium cation, which is the part that binds to the phosphate group. This binding alters the local flexibility of the DNA backbone, which is carried by the covalently-bonded chain of deoxyribose sugars and phosphate groups. Any model compound that is representative of this flexibility should, at a minimum, account for all nearest-neighbor interactions between nucleotide backbone units. One structure that meets these requirements is a combination of three deoxyribose sugars, with two intermediate phosphate groups, as well as one central nucleobase that incorporates the most dominant sources of sequence-dependent motions. Additional non-local interactions beyond neighboring nucleotides, while present, are likely to be less influential to DNA elasticity. They are beyond the scope of this study, and are a subject for future investigation.

\subsection{Specifying the Configuration of the Model Complex}
With the model complex selected, the question turns to determining an appropriate set of variables specifying its atomic coordinates. Such information is necessary both for determining average molecular configurations, and for characterizing the principal modes of fluctuation from this average. There are two parts to the problem: 1) Specifying the configuration of the sugar-phosphate backbone unit, and 2) Specifying the position and orientation of the guanidinium group with respect to the backbone. 

\begin{figure}[h]
\includegraphics[scale=.47]{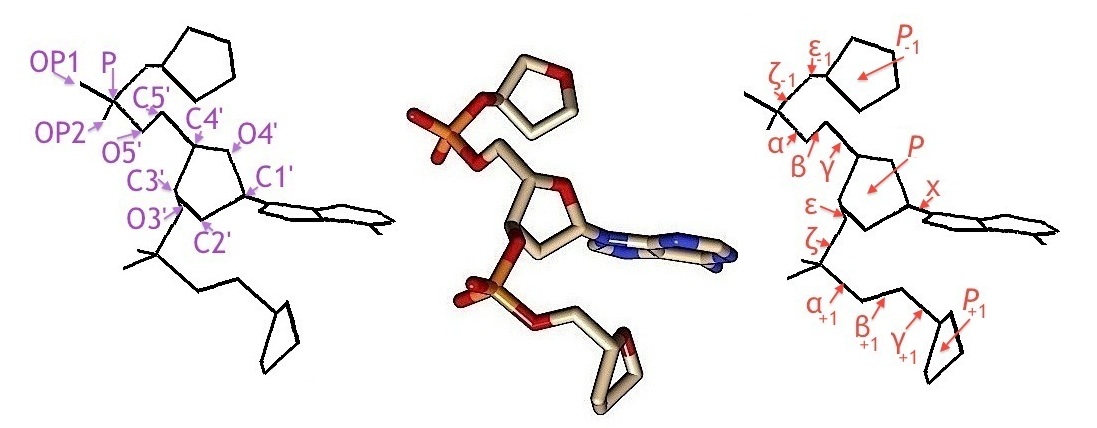} \\
\includegraphics[scale=.75]{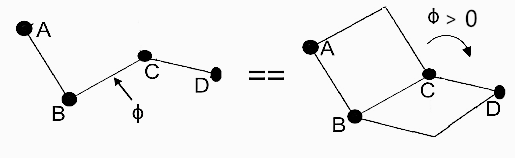}
\caption{The detailed parameters specifying the conformation of the sugar-phosphate backbone in the model complex. (Top) From left to right are a stick image with selected non-hydrogen atoms labeled, an all-atom molecular graphic, and a stick image with the dihedral angles and pseudorotation phase angles labeled. In the all-atom molecular graphic, oxygen is colored red, phosphorus orange, carbon beige, and nitrogen blue, with hydrogens not shown for clarity. (Bottom) Displayed is the chosen positive sign convention for the dihedral angle $\phi$ between four atoms A-B-C-D, defined to be the angle between the planes formed by A-B-C and by B-C-D, with the angle taken to be zero when the atoms are in a planar, \textit{cis} conformation.} 
\label{BackboneCoordinates}
\end{figure}

\subsubsection{Backbone Conformation}

The problem of specifying the backbone coordinates is reduced by the observation that covalent bond lengths in crystal structures are, to a good approximation, fixed at experimentally prescribed values \cite{WKO1996JACS}. Furthermore, except for the covalent linkages formed by the deoxyribose sugars, bond angles are also approximately fixed. The conformation of the deoxyribose sugars, meanwhile, is well described by the phase angle of pseudorotation $\textit{P}$, which specifies the puckering of the furanose ring \cite{Sundaralingam}. With these simplifications, the backbone conformation is specified by the dihedral angles $\alpha$, $\beta$, $\gamma$, $\epsilon$, and $\zeta$ describing covalent bond links between adjacent sugars, the glycosydic torsion angle $\chi$ connecting the central deoxyribose to the nucleobase, and the pseudorotation phase angles $\textit{P}$ of the deoxyribose sugars, as illustrated in Figure \ref{BackboneCoordinates}.  

\subsubsection{Salt-Bridge Configuration}
\begin{figure}[t!]
\includegraphics[scale=.7]{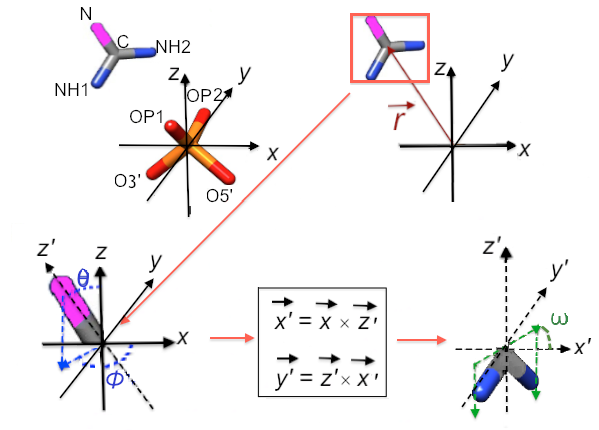}
\caption{Displayed is a schematic of the six variables necessary to represent the configuration of a guanidinium cation with respect to a phosphate group.  Coordinates are chosen so that phosphorus lies at the origin, OP1 and OP2 lie in the $\it{y-z}$ plane at equal and opposite values of $\it{y}$, and O5$^{'}$ lies in the $\it{x-z}$ plane, with positive $x$ and negative $z$. With this choice of coordinate frame, the translational parameters of the guanidinium are specified by the vector $\vec{r}$ describing the displacement of the guanidinium carbon C from the phosphorus atom P. The position of this carbon is then taken to be the origin of a new set of coordinates $x^{'}$, $y^{'}$, and $z^{'}$. These coordinates are defined such that the non-end-group nitrogen N lies on the positive $z^{'}$-axis, the $x^{'}$-axis is set by the cross product of the $x$- and $z^{'}$-axes, and the $y^{'}$-axis is set by the cross product of the $z^{'}$- and $x^{'}$-axes. With this set of coordinates, the rotational degrees of freedom are given by the Euler angles $\theta$ and $\phi$ that the $z^{'}$-axis makes with respect to the $z$-axis, and the angle $\omega$ that the NH1-NH2 vector makes with the $x^{'}$-axis. Images created with Pymol.}
\label{SchematicBridge}
\end{figure}

The specification of the coordinates of the guanidinium is simplified by the observation that its C-N bond angles and bond lengths vary negligibly from 1.33 \AA \ and 120$^{\circ}$, respectively. Thus, the guanidinium cation can be treated as a rigid body with a trigonal planar geometry, and its position and orientation with respect to the backbone reduces to finding three translational and three rotational rigid body parameters, as shown in Figure \ref{SchematicBridge}. Without loss of generality, the phosphorus atom can be defined to be the origin, with the side-group oxygens OP1 and OP2 positioned symmetrically in the $\it{y-z}$ plane. The three translational parameters of the guanidinium can be taken to be the position vector $\vec{r}$ of the central carbon C with respect to the phosphorus. Two angles $\theta$ and $\phi$ then set the orientation of the non-end-group nitrogen N, and an angle $\omega$ describes the remaining rotational freedom of the end-group nitrogens NH1 and NH2.

\section{Methods}
In this section, the procedures for determining the principal components of backbone deformation and primary clusters of guanidinium-phosphate interaction are described. Subsequently, the methodology for electronic structure calculations is expanded upon. Particular focus is given to how these calculations couple to the bioinformatics analyses. 
 
\subsection{Extracting Functional Motions from Crystal Structures}
\subsubsection{Principal Component Analysis of the Sugar-Phosphate Backbone}
Statistical analysis is performed on a non-redundant dataset of protein-bound DNA obtained from the Nucleic Acid Database \cite{BermanOlson92} and reported in a previous publication \cite{Yusufaly13}. From this dataset, a fourteen-parameter data vector is generated that characterizes the atomic configuration of the model complex illustrated in Figure \ref{BackboneCoordinates}. This vector includes a single glycosidic base-sugar torsion angle $\chi$, three sugar pucker phase angles $P$ for each of the three deoxyribose sugars (converted to Cartesian coordinates using an algorithm previously developed by Olson \cite{WKO1982JACS}), and ten dihedral angles along the backbone. The total collection of data vectors is then sorted into four groups based on the identity of the central nucleobase, and each group of data is separately standardized and subjected to principal component analysis. Using a scree test, the four highest eigenvalues, corresponding to dominant modes of deformation, are extracted for each group. 

\subsubsection{Clustering of Guanidinium-Phosphate Salt Bridges}
\begin{figure}[h]
\includegraphics[scale=.31]{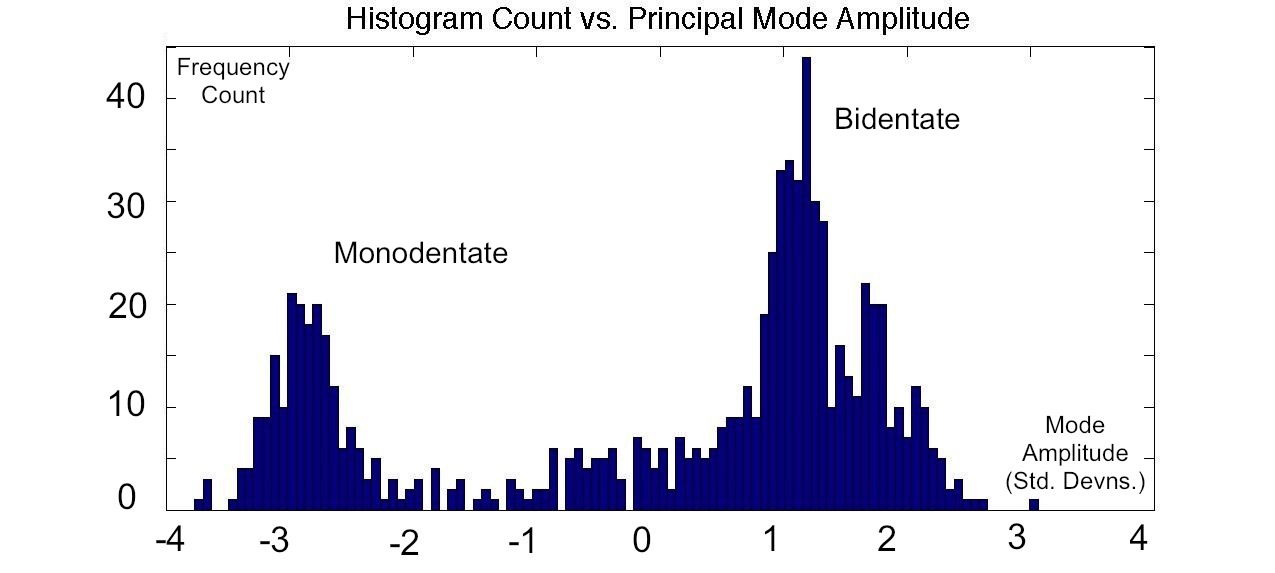} \\
\includegraphics[scale=.5]{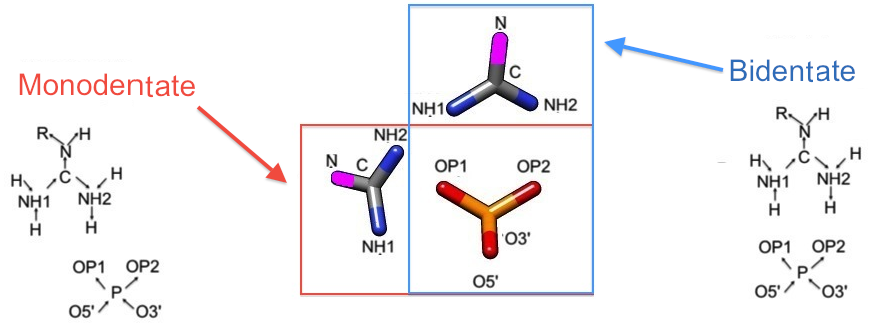}
\caption{Histogram of the frequency count of the amplitude of the dominant principal component, with one hundred equally spaced bins from -4 to 4. The units of measurement are standard deviations from the average value, so that if the average coordinates of the guanidinium are $\langle \vec{q} \rangle = (\langle \vec{r} \rangle, \langle \theta \rangle, \langle \phi \rangle, \langle \omega \rangle)$, and the direction of deformation along the principal component is parallel to the vector $\vec{\lambda}_{parallel} = (\Delta \vec{r}, \Delta \theta, \Delta \phi, \Delta \omega)$, then the PCA vector is normalized to be $\vec{\lambda}_{PCA} = f \vec{\lambda}_{parallel}$, where $f$ is a scaling factor that imposes the condition that the variance of the projection along the PCA equal 1, $\langle |\vec{q} \cdot \vec{\lambda}_{PCA} - \langle \vec{q} \cdot \vec{\lambda}_{PCA} \rangle |^{2} \rangle$. The distribution of the amplitude along this determined principal component $\vec{\lambda}_{PCA}$ is observed to peak around two central clusters, depicted below the histogram:  1) `monodentate' bridges, in which only one hydrogen bond is formed between guanidinium and phosphate, and 2) `bidentate' bridges, in which two hydrogen bonds are formed, as displayed. This justifies, for an initial study, a `mean-field' approximation in which the configuration of the guanidinium cation can be taken as adopting one of two `average' values. These average values are determined by K-means clustering. Histogram created in MATLAB \cite{MATLAB}}.
\label{Hist}
\end{figure}

The analysis of DNA-histone interactions is performed on an ensemble of 83 high-resolution crystal structures of nucleosomal DNA. Within this ensemble, arginine-phosphate contacts are observed to be the most common mode of interaction between the histones and the sugar-phosphate backbone. Thus, an initial dataset is created, consisting of 1556 structural examples in which an arginine nitrogen is less than 4.0 \AA \ away from the phosphorus atom. 

This dataset is further curated so that it only includes structures in which the minimum distance between an end-group nitrogen (NH1 or NH2) and a side-group oxygen (OP1 or OP2) is at least 1.6 \AA \ less than the minimum distance between the non-end-group nitrogen (N) and a side-group oxygen. This step is necessary to remove any `anomalous' structures in which the guanidinium cation may not be hydrogen bonded to the phosphate through the end-group nitrogens. While it is conceivable that arginines may interact with the phosphate in ways different from this, including for example hydrogen bonding of the non-end-group nitrogen to the phosphate, such interactions are beyond the scope of the present analysis, and are a subject for future investigation. As it turns out, the chosen constraints account for over half of all significant arginine-phosphate interactions, resulting in a working dataset of 790 structural examples of guanidinium-phosphate salt bridges.  

From this working dataset, a six-parameter data vector ($\vec{r}$, $\theta$, $\phi$, $\omega$) is generated that characterizes the configuration of a salt bridge, as described in Figure \ref{SchematicBridge}. This collection of data vectors is then standardized and subjected to principal component analysis. A scree test determines that only the first principal component carries a significant fraction of the total variance. Furthermore, a histogram of the frequency distribution of the amplitude of the first principal component, displayed in Figure \ref{Hist}, indicates that the data are localized around two strongly peaked regions:  1) A monodentate cluster, in which only OP1 is hydrogen bonded to an end-group nitrogen, and 2) A bidentate cluster, in which both OP1 and OP2 bond to a separate end-group nitrogen. Because of the sharpness of these peaks, it can be assumed, in a `mean-field' approximation, that the salt bridges only adopt two distinct states corresponding to the centers of each of these clusters. The data are thus sorted by K-means clustering, and the central average of each cluster is taken as one of two possible salt-bridge orientations.

\subsection{Calculating Energy Landscapes with Density Functional Theory}
\begin{figure}[h]
\includegraphics[scale=.45]{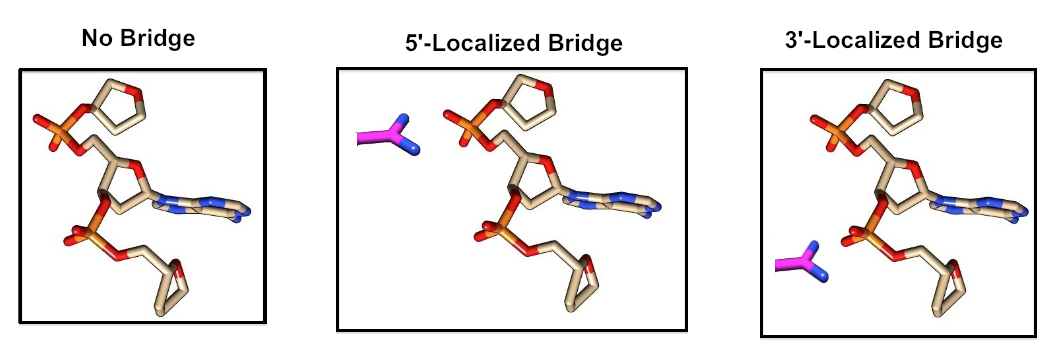} \\
\caption{The energy landscape of each principal component of the backbone is calculated using density functional theory. The energies are first calculated in the absence of any guanidinium cation (left). Calculations are then repeated with salt bridges localized on the 5$^{'}$-phosphate (middle) and 3$^{'}$-phosphate (right). This procedure is repeated for both the monodentate and bidentate configurations determined by K-means clustering, leading to a total of four different salt-bridge environments being simulated. The example salt bridge on the left is of a bidentate form, and the example salt bridge on the right is of a monodentate form. Images created with Pymol. }
\label{BridgePic}
\end{figure}
Having determined both the functional modes of deformation of the backbone, and a representative set of guanidinium-phosphate salt-bridge clusters, the next task is to determine, with vdW-DFT, the elastic energy of deformation of each of the modes in both the absence and presence of different salt bridges. The first step is to sample a series of points along the configurational pathway of each mode, and calculate the energy of each point in the absence of any guanidinium group. From these initial computations, a set of low-energy points along the landscape is determined. Calculations on these low-energy points are then repeated for each of four different types of guanidinium-phosphate salt bridges, namely the set of all combinations of bridges that are localized around the 5$^{'}$ or 3$^{'}$ phosphate group and which lie in either a monodentate or bidentate orientation. 

DFT calculations are performed with the vdW-DF2 functional \cite{Lee10}, as implemented in the Quantum Espresso package \cite{Espresso} via the algorithm developed by Roman-Perez and Soler \cite{VDWAlgorithm}. Standard generalized gradient approximation pseudopotentials \cite{Pseudopotentials} are employed, with a kinetic energy cutoff of 60 Ry (1 Ry = 313.755 kcal/mol). SCF diagonalizations are performed with a convergence criteria of $10^{-6}$ Ry. To ensure efficient convergence of the energy in the presence of the net charges of the phosphate ions, a Makov-Payne electrostatic correction term \cite{MakovPayne} is added. Spurious interaction between artificial periodic images is reduced by placing the system in a cubic supercell of side length 36 Bohr (1 Bohr = 0.529 \AA).

\section{Results and Discussion}
The two main results of this work are that: 1) the configurational fluctuations of the sugar-phosphate backbone, as represented by the dominant principal components, display sequence specificity, and 2) the guanidinium cations interact with the sugar-phosphate backbone to tunably `freeze in' specific backbone deformations. This section begins by discussing the molecular character of the principal components, paying particular attention to signatures of sequence specificity. This is then followed by a presentation of the insights gleaned from DFT calculations, in particular, the observation that the main effect of the guanidinium cations is to apply an approximately linear mechanical stress to the backbone. This stress displays an intricate dependence on many different tunable `knobs', including the chemical identity of the central nucleobase, the choice of phosphate on which the guanidinium cation is localized, and the number of hydrogen bonds that the guanidinium makes with the phosphate. These effects have direct implications for the robust and adaptive control of nucleosome positioning. 

\subsection{Backbone Motions: Bending Virtual Bonds}
\begin{figure}[t]
\includegraphics[scale=.5]{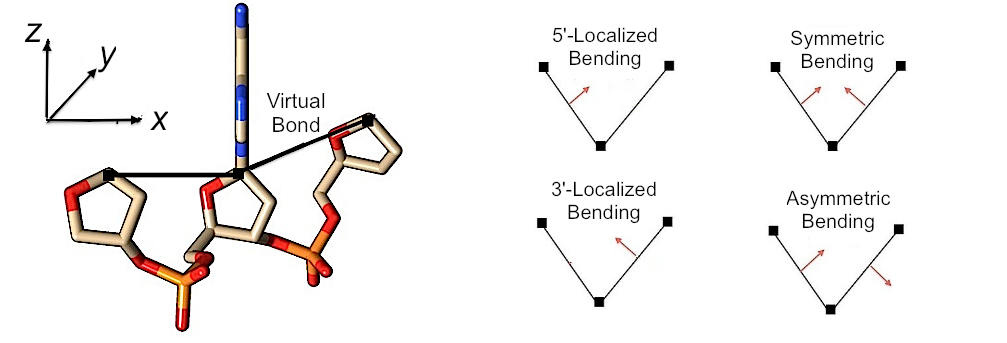}
\caption{(Left) The motions of the sugar-phosphate backbone unit can be simplified by using a reduced description in terms of `virtual bonds' between the C1$^{'}$ atoms on each of the deoxyribose sugars. Then, the complicated collection of atoms in the nucleotide is reduced to a simple virtual triatomic `molecule'. Image created with Pymol. (Right) The deformations of a linear triatomic molecule can be described in terms of the relative motions of each of the two bonds. }
\label{SchematicPolymer}
\end{figure}

\begin{figure}[t]
\includegraphics[scale=.25]{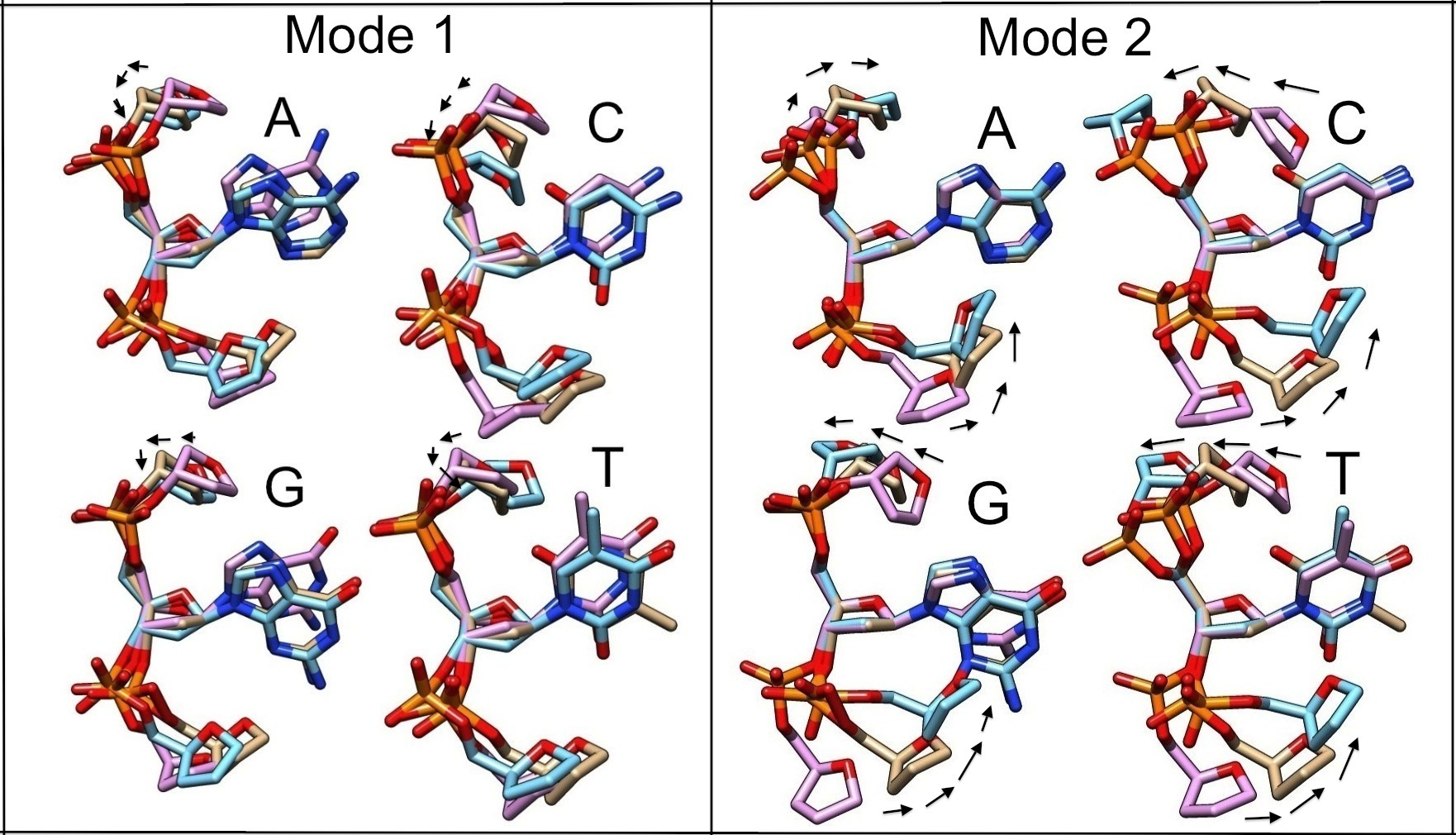}
\caption{Displayed here are the first and second principal modes of deformation, for each of the four different nucleobases. Images are superimposed such that the C1$^{'}$, C3$^{'}$ and C4$^{'}$ atoms of the central deoxyribose sugar are fixed in position. Any bending motions are accompanied by black arrows guiding the direction of motion. The molecular images are color coded such that the beige carbon colored units are associated with the average backbone conformation, and the pink and blue carbon colored units are associated with -1 and 1 standard deviations of deformation away from the average, respectively. Images created with Pymol.}
\label{Modes1and2}
\end{figure}

\begin{figure}[t]
\includegraphics[scale=.25]{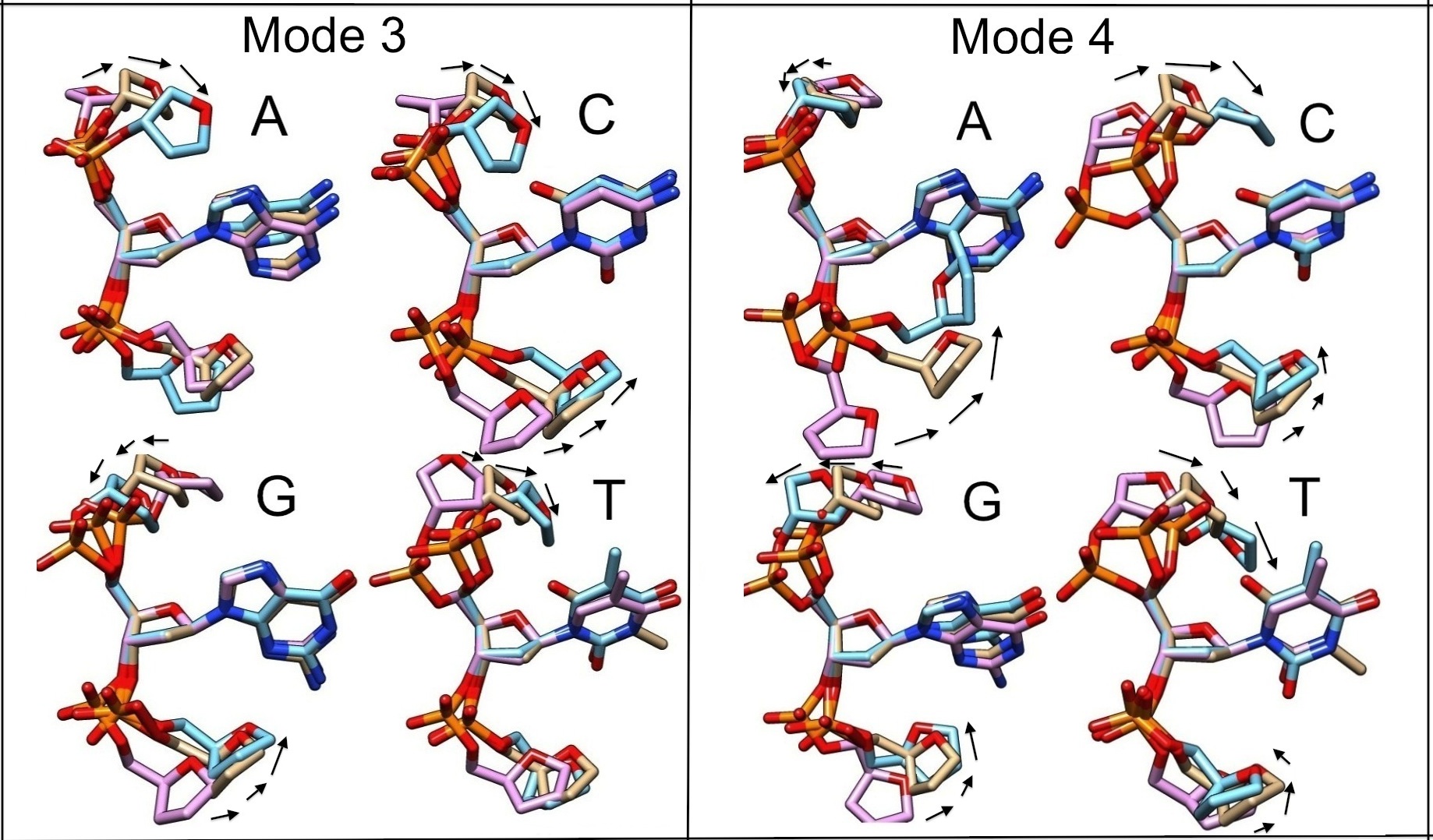}
\caption{Displayed here are the third and fourth principal modes of deformation, for each of the four different nucleobases. For detailed annotation, see the caption of Figure \ref{Modes1and2}.} 
\label{Modes3and4}
\end{figure}

Full results regarding the quantitative coefficients and fractions of total variance captured by each of the four highest principal components are presented in the Supplementary Information. Here, the focus shall be on developing an intuition regarding the qualitative character of these principal modes. In order to develop this intuition, it is useful to switch from the all-atom picture of the backbone to a more coarse-grained view, in which the covalent chain connecting the C1$^{'}$ atoms on adjacent sugars is represented as a `virtual bond' \cite{Flory, OlsonVirtualBond}, as illustrated in Figure \ref{SchematicPolymer}. However, it must be stressed that this coarse-grained description cannot serve as a replacement for a complete quantitative description in terms of the original dihedral angles and sugar puckers. The virtual bond is, again, simply a qualitative heuristic that serves to provide a bird's-eye view of some particularly striking features of the deformation, as a description in terms of a collection of microscopic parameters does not by itself provide particularly useful insights. A deeper analysis and interpretation of the functional roles of the deformations in terms of the detailed values of the microscopic parameters is beyond the scope of this work, but is a topic for future research.

The virtual bond perspective allows the motion of the backbone to be expressed in terms of deformations of a virtual triatomic `molecule', analogous to the well-studied IR vibrations of more well-known triatomic molecules such as H$_{2}$O \cite{Atkins}. For a particular principal component, each virtual bond can be viewed as either increasing, decreasing, or negligibly changing the angle that it makes with respect to the central nucleobase. Then, to a first approximation, a particular combination of bond motions can be characterized as being in one of four classes: 1) 5$^{'}$-Localized bending, in which only the 5$^{'}$-end sugar appreciably moves; 2) Symmetric bending, in which the two bonds move `in-phase'; 3) 3$^{'}$-Localized bending, in which only the 3$^{'}$-end sugar appreciably moves; 4) Asymmetric bending, in which the two bonds move `out-of-phase'. A schematic of the various bending combinations is displayed in Figure \ref{SchematicPolymer}.

The principal components are observed to display a complex dependence on the chemical identity of the central nucleobase. In spite of this, some general patterns and trends do emerge, as displayed in Figures \ref{Modes1and2} and \ref{Modes3and4}. The principal component with the highest amount of the total variance, hereafter labeled the first principal component, displays the least amount of qualitative sequence dependence, adopting a 5$^{'}$-localized bending motion in which the 3$^{'}$-end sugar merely rotates in position. 

Sequence behavior becomes much more diverse for the second, third and fourth components. The second principal component is observed to take the form of asymmetric bending for cytosine, guanine and thymine, but adopts a symmetric bending for adenine. The third principal component demonstrates adenine and thymine performing 5$^{'}$-localized bending, cytosine symmetrically bending, and guanine asymmetrically bending. And finally, the fourth principal component displays a dependence on purine vs. pyrimidine character, being an asymmetric bend for adenine and guanine but a symmetric bend for cytosine and thymine. While these classifications are only qualitative heuristics, they serve to demonstrate the point that the fluctuations of the sugar-phosphate backbone encode sequence information.

\subsection{Tuning Energy Landscapes via Adjustment of Salt Bridges}
The full results of energy vs. mode amplitude for each of the different modes and salt-bridge configurations are relegated to the Supplementary Information. The main text focuses on extracting the energetic effect of the salt bridges, in particular the generation of an approximately linear mechanical stress signal that couples to each of the principal components in a sequence-specific manner. 

\begin{figure}[h]
\includegraphics[scale=.42]{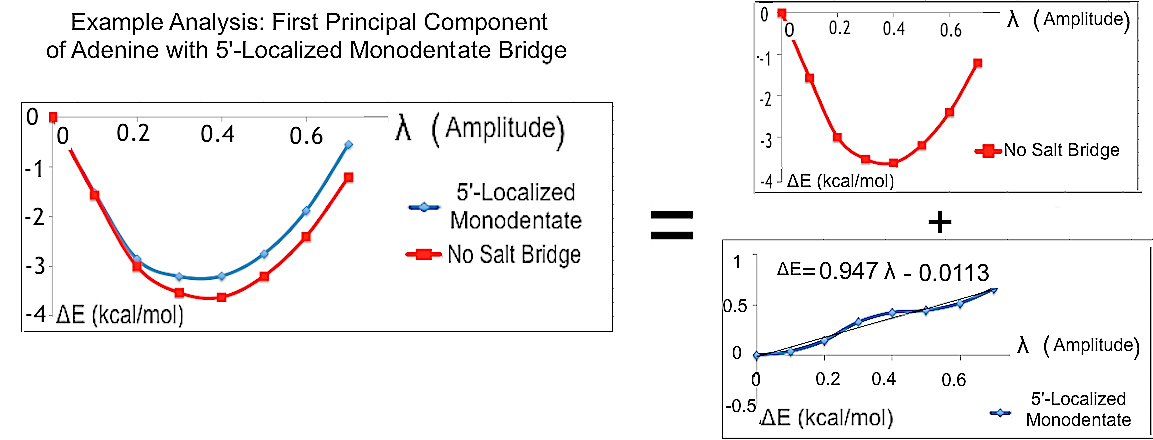}
\caption{Displayed here is an example of the procedure used to extract the mechanochemical stress $\sigma_{\lambda}$ induced on a particular principal component $\lambda$ by a particular type of salt-bridge configuration. In this case, the illustration is provided by the 5$^{'}$-localized monodentate bridge on the first principal component of adenine. The energy landscapes along the mode are computed both with and without the salt bridge, and plots are standardized so that the point of zero mode amplitude is the zero-point reference energy. This allows the energy landscape of the mode in the presence of the salt bridge to be decomposed into the sum of the landscape in the absence of the salt bridge and an approximately linear component representative of the effects of the salt bridge. This component is least-squares fit to a line, and the resulting slope approximates $\sigma_{\lambda}$.This procedure can then be repeated for each of the other three different salt-bridge configurations, and then further repeated for all the different principal components and nucleobases. Physically, this linear mechanochemical stress shifts the equilibrium amplitude of the principal mode. In particular, if the reference state energy is $E_{0}(\lambda) = \frac{1}{2} k (\lambda - \lambda_{0})^{2}$, where $k$ and $\lambda_{0}$ are respectively the spring constant and original equilibrium position, the presence of the mechanochemical stress $\sigma_{\lambda}$ shifts the energy to be $E(\lambda) = \frac{1}{2} k (\lambda - \lambda_{0})^{2} + \sigma_{\lambda} \lambda = \frac{1}{2} k (\lambda - \lambda_{0} - \frac{\sigma_{\lambda}}{k})^{2} +$ a constant term that has no physical effect on the energies and forces. Thus, the equilibrium position has been shifted from $\lambda_{0}$ to $\lambda_{0} - \frac{\sigma_{\lambda}}{k}$, and so a higher amplitude mechanochemical stress corresponds to a larger shift in mode amplitude.}
\label{exampleAPCA1}
\end{figure}

\begin{figure}[h]
\includegraphics[scale=.32]{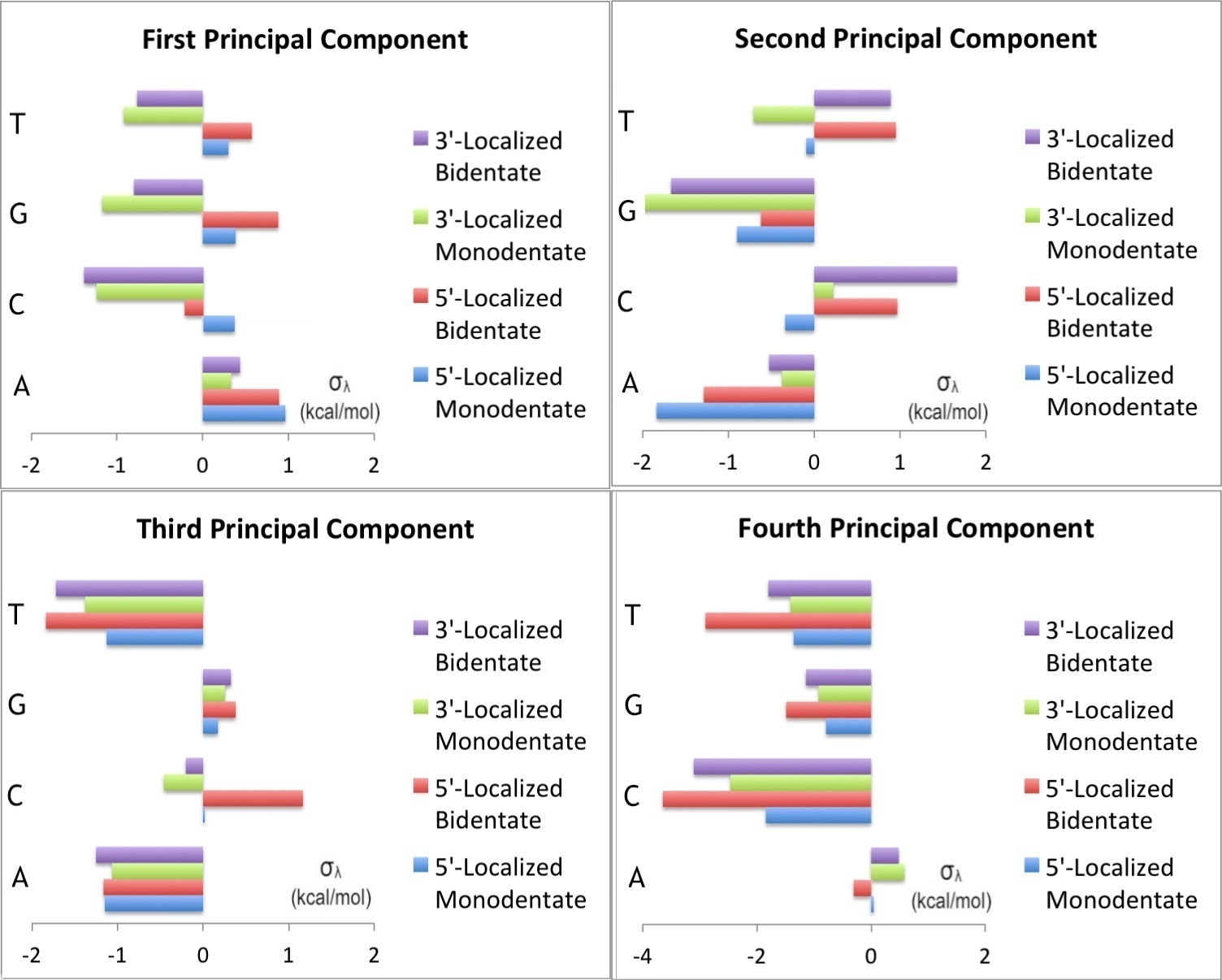}
\caption{Presented here are the results for the different mechanochemical stresses $\sigma_{\lambda}$ for each of the four principal components in the presence of different salt-bridge forms and different central nucleobases. The $x$-axis displays mechanochemical stresses, with units of kcal/mol resulting from the fact that mechanochemical stresses are defined as changes in energy over change in unitless principal mode amplitude.}
\label{Forces}
\end{figure}

To extract this signal, plots of energy vs. mode amplitude are generated for each of the modes in both the presence and absence of salt bridges. From these plots, the energy landscapes of salt-bridged modes are decomposed into the sum of a reference landscape with no salt bridge present and a perturbation that reflects the elastic energy contribution arising from the presence of the guanidinium group. This perturbation contribution is then least-squares fit to a linear function, $\Delta E(\lambda) = \sigma_{\lambda} \ \lambda$, where $\lambda$ is the amplitude of the principal component in units of standard deviations from the mean. The resulting coefficient $\sigma_{\lambda}$ is the linear mechanochemical stress along the axis of deformation of the principal component. An illustration of the procedure is given in Figure \ref{exampleAPCA1} and a full display of the resulting stresses $\sigma_{\lambda}$ is shown in Figure \ref{Forces}.

As seen in Figure \ref{Forces}, the salt-bridge induced stresses display a complex multi-pronged dependence on base sequence, salt-bridge denticity, and phosphate positioning of the guanidinium group. Even for the first principal component, in which the atomic deformation is a 5$^{'}$-localized bending irrespective of base identity, the nature of the salt-bridge induced stresses and their dependence on denticity and positioning differs for adenine as compared to cytosine, guanine, and thymine. 

These effects continue to hold true for other groups of similar deformations. For the second principal component, in which cytosine, guanine, and thymine all asymmetrically bend, the coupling of mechanical stress to salt-bridge denticity and positioning is different for the purine guanine compared to the pyrimidines cytosine and thymine. The third principal component, which groups adenine and thymine together as 5$^{'}$-localized bends, shows that the mechanical effect of the salt bridge on adenine is slightly weaker than it is on thymine.

\subsection{Discussion}
\begin{figure}[h]
\includegraphics[scale=.55]{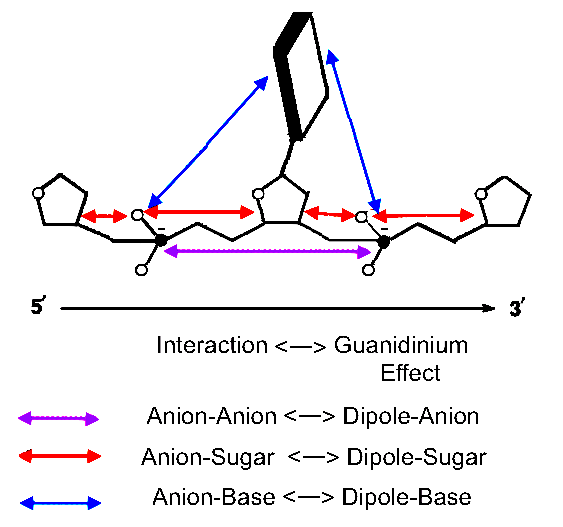}
\caption{The anionic phosphate groups non-covalently interact with each other, the deoxyribose sugars, and the aromatic nucleobases. The combined effect of these nonlocal forces is an `intermolecular' stress arising near the phosphate group. When a guanidinium cation neutralizes one of the phosphate groups, it also modifies these non-covalent interactions and the resulting intermolecular stress. Modification of denticity and positioning further adjust the character of the non-covalent interactions, allowing for a diverse array of tunable `knobs' that induce particular kinds of mechanical deformation.}
\label{NonLocal}
\end{figure}

Chemically, the stabilization of various principal components arises from a combination of `intermolecular', non-covalent effects. The anionic phosphate groups interact electrostatically with each other, and through ion-pi interactions with the aromatic ring sugars and nucleobases. When a guanidinium cation is present, the anionic phosphate group has been reduced to a dipole, and the tuning of denticity tunes the dipole magnitude and orientation. The electrostatic ion-ion interactions of the phosphate groups are reduced to ion-dipole interactions, and the interactions with the aromatic sugars and nucleobases are likewise modified to have a larger contribution from non-electrostatic London dispersion forces. 
 
The result is an effective mechanical load arising from the complex interplay of these different noncovalent interactions, and it is this mechanical load that causes the linear mechanochemical stress which activates specific combinations of principal component deformations. Altogether, the combination of sequence, salt-bridge positioning, and denticity serves as a collection of tunable `knobs' that histones can use to locally activate particular combinations of backbone deformations.

\subsubsection*{Implications for Nucleosome Positioning}

From the point of view of nucleosomes, one of the most interesting consequences of the salt bridges is their modulation of the helical periodicity of the DNA backbone. In canonical DNA forms, such as B-DNA or undertwisted A-DNA, the backbone torsion angles display a consistent periodicity commensurate with the spacing between adjacent base-pairs. In other words, the torsion angles $\alpha$, $\beta$, $\gamma$, $\epsilon$, and $\zeta$ are equal to $\alpha+1$, $\beta+1$, $\gamma+1$, $\epsilon+1$, and $\zeta+1$, respectively. However, the principal modes of deformation do not necessarily obey this periodicity, as seen most notably in modes that tend toward 5$^{'}$-localized and asymmetric bending type character. As a result, the histones effectively apply an elastic modulating signal to the DNA, arising from the collection of guanidinum-phosphate salt-bridge contacts within the nucleosome. By tuning these local sites of DNA-histone binding, the shape and size of the modulating signal can be controlled. In turn, this size and shape alter the equilibrium positioning of various mechanical deformations, such as the wrapped pathways characteristic of nucleosomes.

A further remarkable feature of biological evolution is the high degree of precision with which these delicate elastic modulations are controlled. The sensitivity of DNA deformations to multiple different variables endows the chromatin with a tremendous amount of adaptability, which enables it to maintain the homeostatic stabilization of nucleosome positions under a diverse set of possible environmental perturbations. At the same time, however, this substantial sensitivity creates an equally substantial challenge concerning the maintenance of robust control. A greater set of sensitive variables for adaptation also means there is a greater set of variables that need to be tightly regulated to maintain a normal biological stasis.

This significance of evolution in determining nucleosome positioning has been increasingly recognized over the past few decades. It has been suggested that there is a genomic code for nucleosome positioning \cite{Segal, Cherstvy2}, with evolutionary conservation of high-affinity nucleosome binding sequences. Additionally, it has been further realized that these sites of high-affinity nucleosome binding tend to repeat themselves at well defined 10 base pair periodicity as opposed to being randomly distributed, a phenomenon known as nucleosome phasing \cite{Struhl, Lohr}.

\begin{figure}[h]
\centering
\includegraphics[scale=.35]{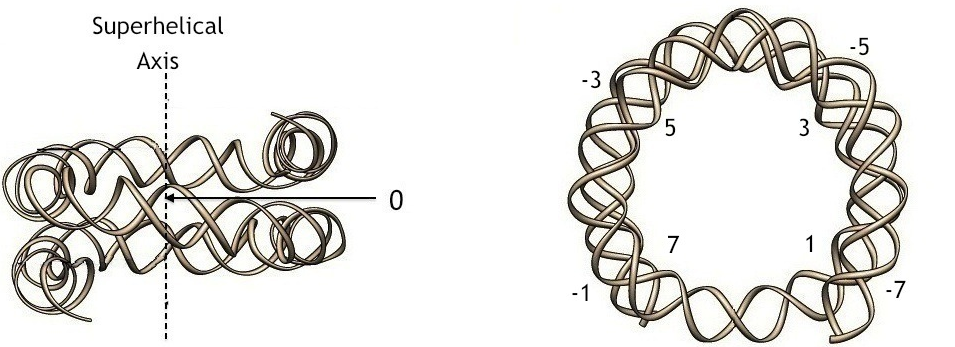}
\includegraphics[scale=.42]{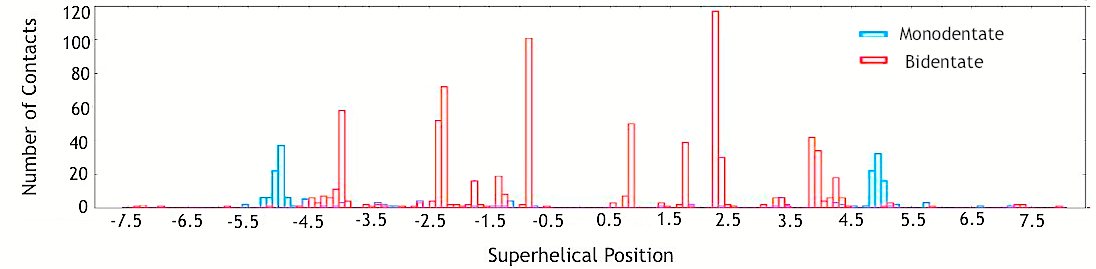}
\caption{(Top) An illustration of the shape of the DNA sugar-phosphate backbone for a high-resolution nucleosomal crystal structure\cite{Luger1}, PDB ID 1kx5, displayed from both a side view and a top-down view. The histones and DNA bases have been removed for clarity. The 147 base pairs of nucleosomal DNA can be viewed as consisting of approximately 15 helical turns of roughly 10 base pairs each, with position along the nucleosome consequently labeled by these helical positions and ranging from -7.5 to 7.5. With this labeling convention, 0 represents the dyad, or the midpoint of the nucleosome that is spatially sandwiched in between the entry and exit points of the nucleosome. (Bottom) A histogram of the frequency of monodentate and bidentate arginine contacts, as a function of helical position, for the 83 nucleosomal crystal structures used in this study. The contacts are observed to localize in well defined clusters.} 
\label{HistoneContacts}
\end{figure}

The results of this study show that a possible evolutionary design principle underlying nucleosome phasing is in the selection for variables that sensitively tune DNA backbone deformations in order to control the nucleosomal wrapping. As the present theoretical calculations show, these deformations are sensitive to both the sequence and positioning of histone-DNA contacts. Furthermore, the diverse nature of these contacts displays a much richer phenomenology than simply electrostatic bindings of cationic amino acids and the anionic phosphate backbone, demonstrating the importance of the relatively underappreciated many body van der Waals interactions in controlling chromatin structure at nanoscopic and mesoscopic length scales. 

Additionally, this work has demonstrated the importance of denticity, a relatively unexplored variable with the potential to be affected by evolution. The number of hydrogen bonds is observed to be of comparable importance to backbone deformation as underlying base sequence and contact positioning. Thus, it is worthwhile to ask if the distribution of such contacts displays similarly non-random behavior characteristic of natural selection. Figure \ref{HistoneContacts} displays a histogram of the nucleosome positions of both monodentate and bidentate contacts in the 83 nucleosomal crystal structures analyzed in this work. As the data show, the positioning of specific types of contacts is far from random, but instead distributed in very well localized clusters. 

Monodentate contacts, in particular, are found to be strongly localized at $\pm$ 5 helical turns with respect to the central nucleosomal dyad. The placement of these contacts is commensurate with regions of the nucleosome that previous researchers \cite{Polach, LiWidom, Tims, Stud1, Stud2} have associated with a high affinity for `invasion' by DNA binding proteins, an effect that is important for active nucleosome remodeling. Specifically, the strongest resistance to DNA unzipping, or strand separation, found in single-molecule experiments \cite{LiAndWang}, occurs around the dyad and in the end regions of the nucleosome at DNA sites roughly four to five helical turns from the dyad, precisely the locations of the termini of nucleosomal DNA structures observed to be anchored by monodentate interactions in Figure \ref{HistoneContacts}. The unzipping of the DNA ends from the histone protein core in single-molecule strand-separation experiments may take advantage of these weaker links, and thus reduced stability. 

In essence, the weakened monodentate contacts appear to serve the function of facilitating the precise microscopic mechanisms of nucleosome unwrapping. The interactions of large protein assemblies with nucleosomes induce the dissociation of the H2A/H2B dimers from the upper and lower surfaces of the core of histone proteins. The monodentate contacts between DNA and H2A-H2B may contribute to this behavior. For example, RNA polymerase II displaces nearly 50 base pairs of DNA from the 5$^{'}$-end of the nucleosome (up to site -2.5 in Figure \ref{HistoneContacts}.) \cite{Kulaeva2010}. The monodentate contacts provide the initial barrier to polymerase invasion of the nucleosome, and appear to compete with the transcription-facilitating protein FACT for access to the H2A/H2B dimer \cite{Hsieh2013}.


The present work suggests that the above evolutionary constraints of having a certain number of contacts with reduced stability must be met while also simultaneously maintaining very specific shape and mechanical stress requirements for the nucleosome. This results in evolutionary selection pressure for a very precise spatial distribution of the necessary van der Waals contacts, and thus of denticity, contact positioning, and DNA sequence. Incidentally, this is also consistent with related work indicating that nucleosome structure and function are highly sensitive to histone sequence, and are in fact disrupted by SIN point mutations of histones \cite{Kurumizaka, Flaus, Olson} which could potentially interfere with the van der Waals contacts.

\section{Conclusions}
In summary, this work has presented a novel integration of structural bioinformatics and van der Waals density functional theory to investigate the effects of a major histone-DNA interaction, the formation of salt bridges between guanidinium arginines and the DNA phosphate group, on the deformations of the DNA sugar-phosphate backbone. Guanidinium-phosphate complexes are observed to occur in both bidentate and monodentate salt-bridge configurations. The combined interplay between denticity, chemical identity of nucleobases, and positioning of the guanidinium group creates a rich array of different mechanochemical stress input signals. These equip the histones with a versatile toolkit for the precise stabilization and control of nucleosome positioning, a toolkit that, in addition, is experimentally observed to be very carefully selected for and organized in evolved living organisms. These results suggest that a possible molecular evolutionary force underlying the structure and function of chromatin is in the selection for highly detailed distributions of van der Waals contacts, and thus intermolecular forces controlling DNA architecture. 

Furthermore, the conclusions derived from this study are expected to have broader implications for the understanding of protein-DNA interactions in general, beyond simply arginine-phosphate bindings in nucleosomes. Firstly, it must be emphasized that the collection of protein-DNA crystal structures that was used to extract the principal components of deformation of the backbone is a non-redundant ensemble of many different DNA microenvironments, and is not artificially restricted to any one particular biophysical condition. Thus, the principal axes derived from the dataset serve as equally valid estimates that can characterize DNA motions in any \textit{in vivo} environment, whether that be, for example, within the packaging of nucleosomes, in the presence of transcription factors and other regulatory machinery, or while the DNA is undergoing transcription, replication or repair. It is certainly true that in any one specific biological setting, the axes of deformation will be shifted from the coarse-grained averages computed from the entire ensemble; nevertheless, these averages have value as zeroth-order approximations independent of the particular intracellular milieu.

Additionally, previous structural bioinformatics analyses of amino acid-DNA contacts in high-resolution crystal structures \cite{WKO1, WKO2} have highlighted that arginine-phosphate bindings, in addition to being the most common mode of interaction between histones and DNA, are also one of the most common classes of protein-DNA interactions in general. They show up as a common binding motif controlling the stability and functionality of many different protein-DNA complexes. While this work only analyzed these contacts in the context of nucleosomal structures, the biochemical constraints of electrostatics and hydrogen bonding stability suggest that the specific salt-bridge orientations determined here are likely to be fairly universal constraints on the nature of allowed arginine-phosphate interactions in general. As a consequence, the results of this study, highlighting the important functional role of many-body van der Waals effects in controlling the coupling of DNA deformation to arginine orientation, are expected to have a similarly broad range of applicability.

In fact, one can go even further, and point out that the implicated importance of van der Waals dispersion forces is likely to hold true for other protein-DNA interactions beyond simply arginine-phosphate salt-bridges. The structural bioinformatics studies cited above report that the class of observed protein-DNA interactions, and their sequence preferences, are generally constrained to lie within a fairly narrow range of all possibilities. These protein-DNA interactions act primarily through modification of DNA shape and flexibility. In this study, the mechanochemical effects of London dispersion forces on DNA were found to be non-negligible for arginine, a charged amino acid. There is, \textit{a priori}, no reason to expect that similarly important contributions may not also be found for other charged amino acids. And for non-charged amino acids, the relative importance of dispersion, if anything, should increase. 

Therefore, when considering the question of the evolutionary selection for specific protein-DNA interactions, with very particular force transductions and mechanical responses, it is highly questionable to assume that one can get correct answers while neglecting London dispersion forces, particularly many-body effects. In turn, these errors at the microscopic scales can potentially amplify at larger, mesoscopic scales, resulting in quantitatively, and possibly even qualitatively, unreliable models of DNA mechanics and its role in gene regulatory networks. The advent of modern density functional theory promises to be a crucial step in remedying this potential roadblock, and the present work has presented a first step towards its application in the accurate theoretical modeling of protein-DNA interactions.

\section*{Supplementary Information}
The supplementary information presents: 1) the raw numerical output results of the principal component analysis, 2) the raw output of the DFT energy calculations, 3) an explanation of the procedure used to convert pseudorotation phase angles of the deoxyribose sugar into Cartesian coordinates, and 4) details on salt-bridge clustering, with references to original literature for the 83 nucleosomal crystal structures, pairwise sequence alignment of the nucleosome crystal structures and annotated output tables of monodentate and bidentate bridges.

\begin{acknowledgments}
We thank D. Vanderbilt and G. Bowman for valuable discussions. T.Y. thanks Bell Laboratories for financial support through the Lucent Fellowship. This work was partially funded by the U.S.P.H.S. through GM 34809.
\end{acknowledgments}

\linespread{1.5}

\pagebreak 


\begin{thebibliography}{99}

\bibitem{WatsonCrick}
J.D. Watson and F.W. Crick, Nature {\bf 171}, 737-738 (1953).

\bibitem{Turner07}
B.M. Turner, Nature Cell Biol. {\bf 9}, 2-6 (2007).

\bibitem{ChenDent2014}
T. Chen and S.Y.R. Dent, Nature Rev. Gen. {\bf 15}, 93-106 (2014).

\bibitem{Bustamante}
C. Bustamante, Z. Bryant and S.B. Smith, Nature {\bf 421}, 423-427 (2003).

\bibitem{MRS2006}
J. Hafner, C. Wolverton and G. Ceder, MRS Bulletin {\bf 31}, 659-668 (2006).

\bibitem{HK64}
P. Hohenberg and W. Kohn, Phys. Rev. {\bf 136}, B864-B871 (1964).

\bibitem{KS65}
W. Kohn and L.J. Sham, Phys. Rev. {\bf 140}, A1133-A1138 (1965). 

\bibitem{Dion04}
M. Dion, H. Rydberg, E. Schr\"{o}der, D.C. Langreth, B.I. Lundqvist, Phys. Rev. Lett. {\bf 92}, 246401 (2004).

\bibitem{Lee10}
K. Lee, E.D. Murray, L. Kong, B.I. Lungdqvist, D.C. Langreth, Phys. Rev. B. {\bf 82}, 081101(R) (2010). 

\bibitem{Cooper08}
V.R. Cooper, T. Thonhauser, A. Pudzer, E. Schr\"{o}der, B.I. Lundqvist, D.C. Langreth, J. Am. Chem. Soc. {\bf 130}, 1304-1308 (2008).

\bibitem{Yusufaly13}
T. Yusufaly, Y. Li and W.K. Olson, J. Phys. Chem. B {\bf 117(51)}, 16436-16442 (2013).

\bibitem{BirdWolfe99}
A.P. Bird and A.P. Wolffe, Cell {\bf 99}, 451-454 (1999). 

\bibitem{Clauvelin1}
O.I. Kuleaeva, G. Zheng, Y.S. Polikanov, A.V. Colasanti, N. Clauvelin, S. Mukhopadhyay, A.M. Sengupta, V.M. Studitsky and W.K. Olson, J. Biol. Chem. {\bf 287} (24): 20248-20257 (2012).

\bibitem{Clauvelin2}
W.K. Olson, N. Clauvelin, A.V. Colasanti, G. Singh and G. Zheng, Biophys. Rev. {\bf 4}, 171-178 (2012).

\bibitem{Luger1}
K. Luger, A.W. Mader, R.K. Richmond, D.F. Sargent, and T.J. Richmond, Nature {\bf 389} (6648): 251-260 (1997).

\bibitem{Luger2}
C.A. Davey, D.F. Sargent, K. Luger, A.W. Maeder, and T.J. Richmond, J. Mol. Biol. {\bf 319} (5): 1097-1113 (2002).

\bibitem{Mirzabekov}
A.D. Mirzabekov and A. Rich., Proc. Natl. Acad. Sci. {\bf 76} (3): 1118-1121 (1979).

\bibitem{StraussReview}
L.D. Williams and L.J. Maher, Annu. Rev. Biophys. Biomol. Struct. {\bf 29}, 497-521 (2000).

\bibitem{Cherstvy1}
A.G. Cherstvy, J. Phys. Chem. B {\bf 113(13)}, 4242Ð4247 (2009).

\bibitem{Marcovitz}
A. Marcovitz and Y. Levy, Proc. Natl. Acad. Sci. {\bf 108}, 17957-17962 (2011).

\bibitem{Honig}
R. Rohs, S.M. West, A. Sosinsky, P. Liu, R.S. Mann and B. Honig, Nature {\bf 461}, 1248-1253 (2009).

\bibitem{StructBook}
W. Saenger, {\it Principles of Nucleic Acid Structure} (Springer-Verlag, New York, 1984).

\bibitem{Pymol}
The PyMOL Molecular Graphics System, Version 1.5.0.4 Schršdinger, LLC.

\bibitem{Sponer1}
J. Sponer, A. Mladek, J.E. Sponer, D. Svozil, M. Zgarbova, P. Banas, P. Jurecka and M. Otyepka, Phys. Chem. Chem. Phys. {\bf 14}, 15257-15277 (2012).

\bibitem{Sponer2}
A. Mladek, M. Krepl, D. Svozil, P. Cech, M. Otyepka, P. Banas, M. Zgarbova, P. Jurecka and J. Sponer, Phys. Chem. Chem. Phys. {\bf 15}, 7295-7310 (2013).

\bibitem{Frigyesa}
D. Frigyesa, F. Alber, S. Pongor and P. Carloni, J. Mol. Struc. (Theochem), {\bf 574}, 39-45 (2001).

\bibitem{Yang08}
L. Yang, G. Song, A. Carriquiry and R. Jernigan, {\bf 16} (2): 321-330 (2008).

\bibitem{Hub09}
J.S. Hub and B.L. de Groot, PLoS Comput. Biol. {\bf 5} (8): e1000480 (2009).

\bibitem{WKO1996JACS}
A. Gelbin, B. Schneider, L. Clowney, S.H. Hsieh, W.K. Olson and H.M. Berman, J. Am. Chem. Soc. {\bf 118} (3): 519-529 (1996).

\bibitem{Sundaralingam}
C. Altona and M. Sundaralingam, J. Am. Chem. Soc. {\bf 94} (23): 8205-8212 (1972).

\bibitem{BermanOlson92}
H. M. Berman, W. K. Olson, D.L. Beveridge, J. Westbrook, A. Gelbin, T. Demeny, S.H. Hsieh, A.R. Srinivasan and B. Schneider, Biophys. J., {\bf 63}, 751-759 (1992).

\bibitem{WKO1982JACS}
W.K. Olson, J. Am. Chem. Soc. {\bf 104} (1): 278-286 (1982). 

\bibitem{MATLAB}
MATLAB version 7.10.0. Natick, Massachusetts: The MathWorks Inc., 2010.

\bibitem{Espresso}
P. Giannozzi, S. Baroni, N. Bonini, M. Calandra, R. Car, C. Cavazzoni, C Ceresoli, G.L. Chiarotti, M. Cococcioni, I. Dabo, et al. J. Phys.: Cond. Mat., {\bf 21}, 395502 (2009).

\bibitem{VDWAlgorithm}
G. Roman-Perez and J.M. Soler, Phys. Rev. Lett., {\bf 103}, 096102 (2009).

\bibitem{Pseudopotentials}
N. Troullier and J. Martins, Phys. Rev. B., {\bf 43}, 1993-2006 (1991).

\bibitem{MakovPayne}
G. Makov and M.C. Payne, Phys. Rev. B., {\bf 51}, 4014 (1995).

\bibitem{Flory}
P. Flory, {\it Statistical Mechanics of Chain Molecules} (Interscience Publishers, New York, 1969).

\bibitem{OlsonVirtualBond}
W.K. Olson, Macromolecules, {\bf 8} (3): 272-275 (1975).

\bibitem{Atkins}
P.W. Atkins and R.S. Friedman, {\it Molecular Quantum Mechanics} (Oxford University Press, New York, 1997). 

\bibitem{Segal}
E. Segal, Y. Fonduffe-Mittendorf, L. Chen, A. Thastrom, Y. Field, I.K. Moore, J.P. Wang and J. Widom, Nature, {\bf  442}, 772-778 (2006).

\bibitem{Cherstvy2}
D.A. Beshnova, A.G. Cherstvy, Y. Vainhstein and V.B. Teif, PLoS Comput. Biol., {\bf 10} (7),  e1003698 (2014).

\bibitem{Struhl}
K. Struhl and E. Segal, Nature Struct. Molec. Biol., {\bf 20}, 267-273 (2013).

\bibitem{Lohr}
D. Lohr, K. Tatchell and K.E. Van Holde, Cell, {\bf 12} (3): 829-836 (1977).

\bibitem{Polach}
K.J. Polach and J. Widom, J. Mol. Biol. {\bf 254}, 130-149 (1995).

\bibitem{LiWidom}
G. Li and J. Widom, Nature Struct. Molec. Biol., {\bf 11} (8): 763-769 (2004).

\bibitem{Stud1}
M.L. Kireeva, W. Walter, V. Tchernajenko, V. Bondarenk, M. Kashlev and V.M. Studitsky, Mol. Cell, {\bf 9} (3): 541-552 (2002).

\bibitem{Stud2}
M.L. Kireeva, B. Hancock, G.H. Cremona, W. Walter, V.M. Studitsky and M. Kashlev, Mol. Cell, {\bf 18} (1): 97-108 (2005).

\bibitem{LiAndWang}
M. Li and M.D. Wang, Methods Enzym., {\bf 513}, 29-58 (2012).

\bibitem{Kulaeva2010}
O.I. Kulaeva, D.A. Gaykalova, N.A. Pestov, V.V. Golovastov, D.G. Vassylev, I. Artsimovitch and V.M. Studitsky, Nat. Struct. Molec. Biol., {\bf 16}, 1272-1278 (2010).

\bibitem{Hsieh2013}
F.K. Hsieh, O.I. Kulaeva, S.S. Patel, P.N. Dyer, K. Luger, D. Reinberg and V.M. Studitsky, Proc. Natl. Acad. Sci. {\bf 110}, 7654-7659 (2013).

\bibitem{Tims}
H.S. Tims, K. Gurunathan, M. Levitus and J. Widom, J. Mol. Biol. {\bf  411} (2): 430-448 (2011).

\bibitem{Kurumizaka}
H. Kurumizaka, A.P. Wolffe, Mol. Cell. Biol. {\bf 17} (12): 6953-6969 (1997).

\bibitem{Flaus}
A. Flaus, C. Rencurel, H. Ferreira, N, Wiechens and T. Owen-Hughes, EMBO J. {\bf 23} (2): 343-353 (2004).

\bibitem{Olson}
F. Xu, A.V. Colasanti, Y. Li and W.K. Olson, Nucleic Acids Res. {\bf 38} (20): 6872-6882 (2010).

\bibitem{WKO1}
W.K. Olson, A.V. Colasanti, Y. Li, W. Ge, G. Zheng and V.B. Zhurkin, Computational Studies of RNA and DNA (eds. J. Sponer and F. Lankas), pg. 235-257 (2006)

\bibitem{WKO2}
A.V. Colasanti, Y. Li, G. Singh, X.J. Lu and W.K. Olson, Biomolecular Forms and Functions (eds. M. Bansal and N. Srinivasan), pg. 230-246 (2013)

\end{thebibliography}
\end{document}